\documentstyle[twoside,fleqn,espcrc2,epsfig]{article}

\newcommand{\agt}{\,\rlap{\lower 3.5 pt \hbox{$\mathchar \sim$}} \raise 1pt
 \hbox {$>$}\,}
\newcommand{\alt}{\,\rlap{\lower 3.5 pt \hbox{$\mathchar \sim$}} \raise 1pt
 \hbox {$<$}\,}

\title{Quarkonium production in deep-inelastic scattering\thanks{%
To appear in {\it Proceedings of the 6th International Symposium on Radiative
Corrections: Application of Quantum Field Theory to Phenomenology (RADCOR
2002)}, Kloster Banz, Germany, September 8--13, 2002.}
}

\author{B.A. Kniehl\address{II. Institut f\"ur Theoretische Physik, 
Universit\"at Hamburg,\\
Luruper Chaussee 149, 22761 Hamburg, Germany}}

\begin{document}

\begin{abstract}
We discuss the inclusive production of $J/\psi$ mesons in deep-inelastic
scattering (DIS) via the electromagnetic, weak neutral, and charged currents
within the factorization formalism of nonrelativistic quantum chromodynamics.
Theoretical predictions are confronted with experimental data of $ep$ and
$\nu N$ DIS taken by the H1 Collaboration at DESY HERA and the CHORUS 
Collaboration at CERN, respectively.
\end{abstract}

\maketitle

\section{Introduction}

Since its discovery in 1974, the $J/\psi$ meson has provided a useful
laboratory for quantitative tests of quantum chromodynamics (QCD) and, in
particular, of the interplay of perturbative and nonperturbative phenomena.
The factorization formalism of nonrelativistic QCD (NRQCD) \cite{bbl} provides
a rigorous theoretical framework for the description of heavy-quarkonium
production and decay.
This formalism implies a separation of short-distance coefficients, which can 
be calculated perturbatively as expansions in the strong-coupling constant
$\alpha_s$, from long-distance matrix elements (MEs), which must be extracted
from experiment.
The relative importance of the latter can be estimated by means of velocity
scaling rules, i.e.\ the MEs are predicted to scale with a definite power of
the heavy-quark ($Q$) velocity $v$ in the limit $v\ll1$.
In this way, the theoretical predictions are organized as double expansions in
$\alpha_s$ and $v$.
A crucial feature of this formalism is that it takes into account the complete
structure of the $Q\overline{Q}$ Fock space, which is spanned by the states
$n={}^{2S+1}L_J^{(c)}$ with definite spin $S$, orbital angular momentum $L$,
total angular momentum $J$, and colour multiplicity $c=1,8$.
The hierarchy of the MEs predicted by the velocity scaling rules is explained
for the $J/\psi$, $\chi_{cJ}$, and $\psi^\prime$ mesons in
Table~\ref{tab:vsr}.
In particular, this formalism predicts the existence of colour-octet (CO)
processes in nature.
This means that $Q\overline{Q}$ pairs are produced at short distances in
CO states and subsequently evolve into physical, colour-singlet (CS) quarkonia
by the nonperturbative emission of soft gluons.
In the limit $v\to0$, the traditional CS model (CSM) \cite{ber} is recovered.
The greatest triumph of this formalism was that it was able to correctly 
describe \cite{ebr} the cross section of inclusive charmonium
hadroproduction measured in $p\overline{p}$ collisions at the Fermilab
Tevatron \cite{abe}, which had turned out to be more than one order of
magnitude in excess of the CSM prediction.
\vspace*{-0.5cm}
\begin{table}[h]
\caption{Values of $k$ in $\left\langle{\cal O}^H[n]\right\rangle\propto v^k$
for $H=J/\psi,\chi_{cJ},\psi^\prime$.}
\label{tab:vsr}
\begin{tabular}{c|cc}
$k$ & $J/\psi$, $\psi^\prime$ & $\chi_{cJ}$ \\
\hline
3 & ${}^3\!S_1^{(1)}$ & --- \\
5 & --- & ${}^3\!P_J^{(1)}$, ${}^3\!S_1^{(8)}$ \\
7 & ${}^1\!S_0^{(8)}$, ${}^3\!S_1^{(8)}$, ${}^3\!P_J^{(8)}$ & --- \\
\end{tabular}
\end{table}
\vspace*{-0.75cm}

In order to convincingly establish the phenomenological significance of the
CO processes, it is indispensable to identify them in other kinds of
high-energy experiments as well.
Studies of charmonium production in $ep$ photoproduction, $ep$ and $\nu N$
deep-inelastic scattering (DIS), $e^+e^-$ annihilation, $\gamma\gamma$
collisions, and $b$-hadron decays may be found in the literature; see
Ref.~\cite{bra} and references cited therein.
Here, $N$ denotes a nucleon.
Furthermore, the polarization of charmonium, which also provides a sensitive
probe of CO processes, was investigated \cite{ben,bkl}.
Until very recently, none of these studies was able to prove or disprove the
NRQCD factorization hypothesis.
However, preliminary data of $\gamma+\gamma\to J/\psi+X$ taken by the DELPHI
Collaboration \cite{delphi} at LEP2 provide first independent evidence for it
\cite{gg}.

In this presentation, we review recent studies of $J/\psi$ inclusive
production in DIS via the electromagnetic \cite{ep}, weak neutral (NC)
\cite{nun}, and charged currents (CC) \cite{cc} to lowest order (LO) in the
NRQCD factorization formalism.

\boldmath
\section{$ep$ DIS via the electromagnetic current}
\unboldmath

\vspace*{-1.cm}
\begin{figure}[h]
\epsfig{figure=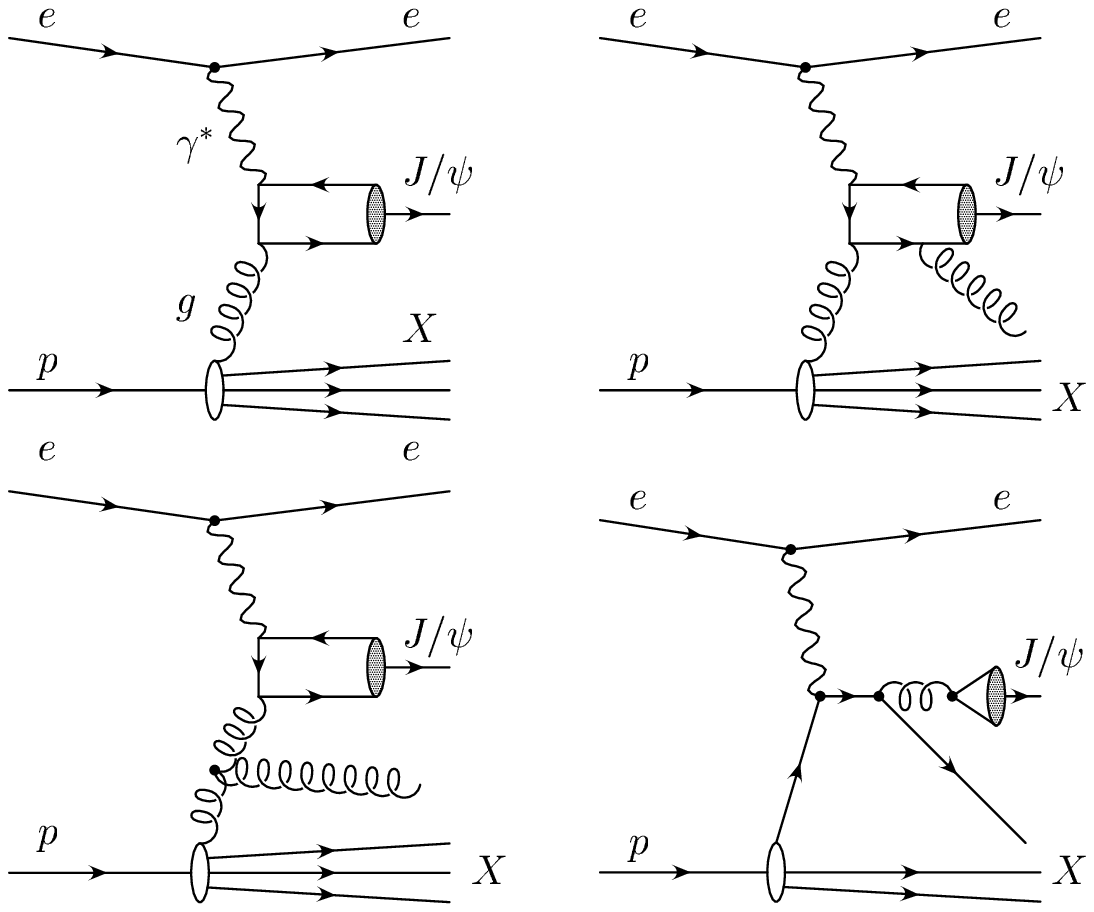,width=7.5cm}
\vspace*{-1.25cm}
\caption{Feynman diagrams for $e+p\to e+J/\psi+X$ in DIS.}
\label{fig:epfey}
\end{figure}
\vspace*{-0.75cm}
The Feynman diagrams for $e+p\to e+J/\psi+X$ in DIS are depicted in 
Fig.~\ref{fig:epfey}.
This process allows for a particularly clean test of the NRQCD factorization
hypothesis, since the large photon virtuality $Q^2$ ensures that perturbative
QCD is applicable and that the resolved-photon contribution, which suffers
from our imperfect knowledge of the parton density functions (PDFs) of the
photon, is greatly suppressed.
Diffractive processes, which cannot yet be reliably described within purely 
perturbative QCD, can be excluded by imposing an upper cutoff on the
elasticity variable $z$, which also eliminates the $\gamma^\star g$ fusion
process (first diagram in Fig.~\ref{fig:epfey}).
The diagrams in the lower panel correspond to CO processes, while the right
one in the upper panel already contributes in the CSM.

\begin{figure*}[t]
\centerline{
\epsfig{figure=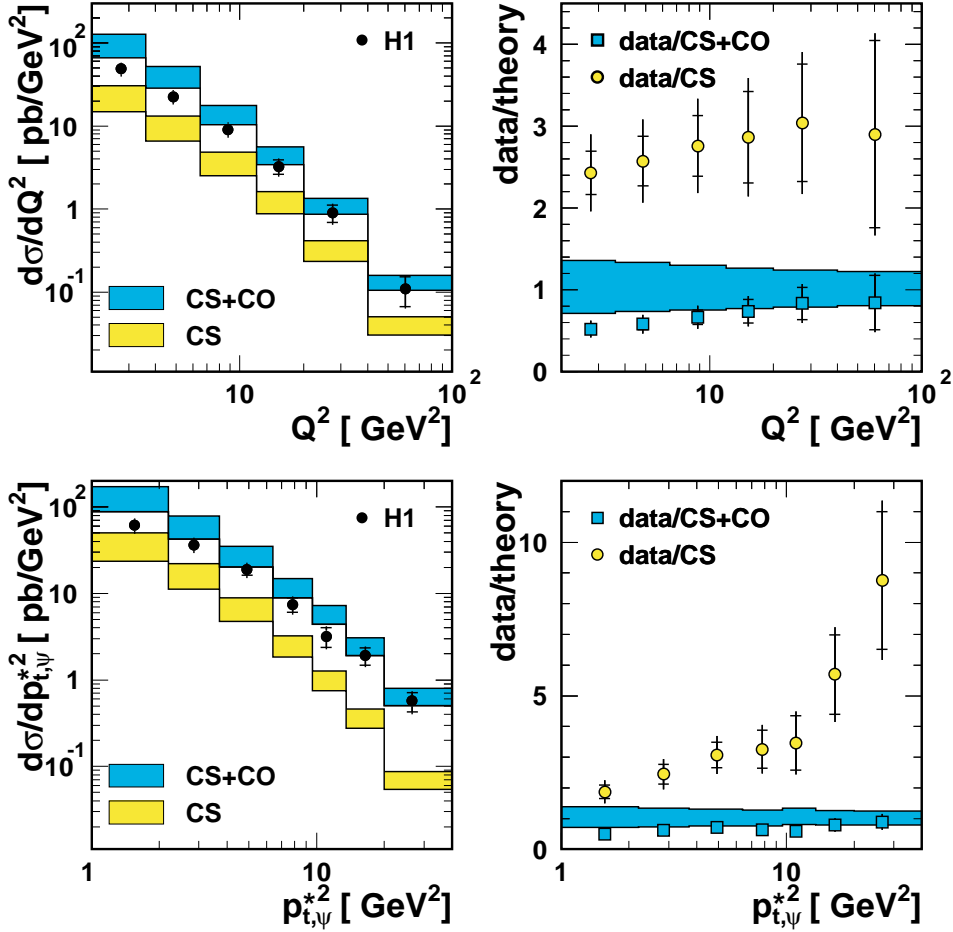,height=12.5cm,bbllx=47pt,bblly=164pt,bburx=542pt,%
bbury=652pt,clip=}
}
\vspace*{-1.cm}
\caption{Comparison of the CSM and NRQCD predictions \protect\cite{ep} for the
$Q^2$ and $p_{t,\psi}^{\star2}$ distributions of $e+p\to e+J/\psi+X$ with H1
data \protect\cite{h1}.}
\label{fig:h1}
\end{figure*}

\begin{figure*}[t]
\centerline{
\epsfig{figure=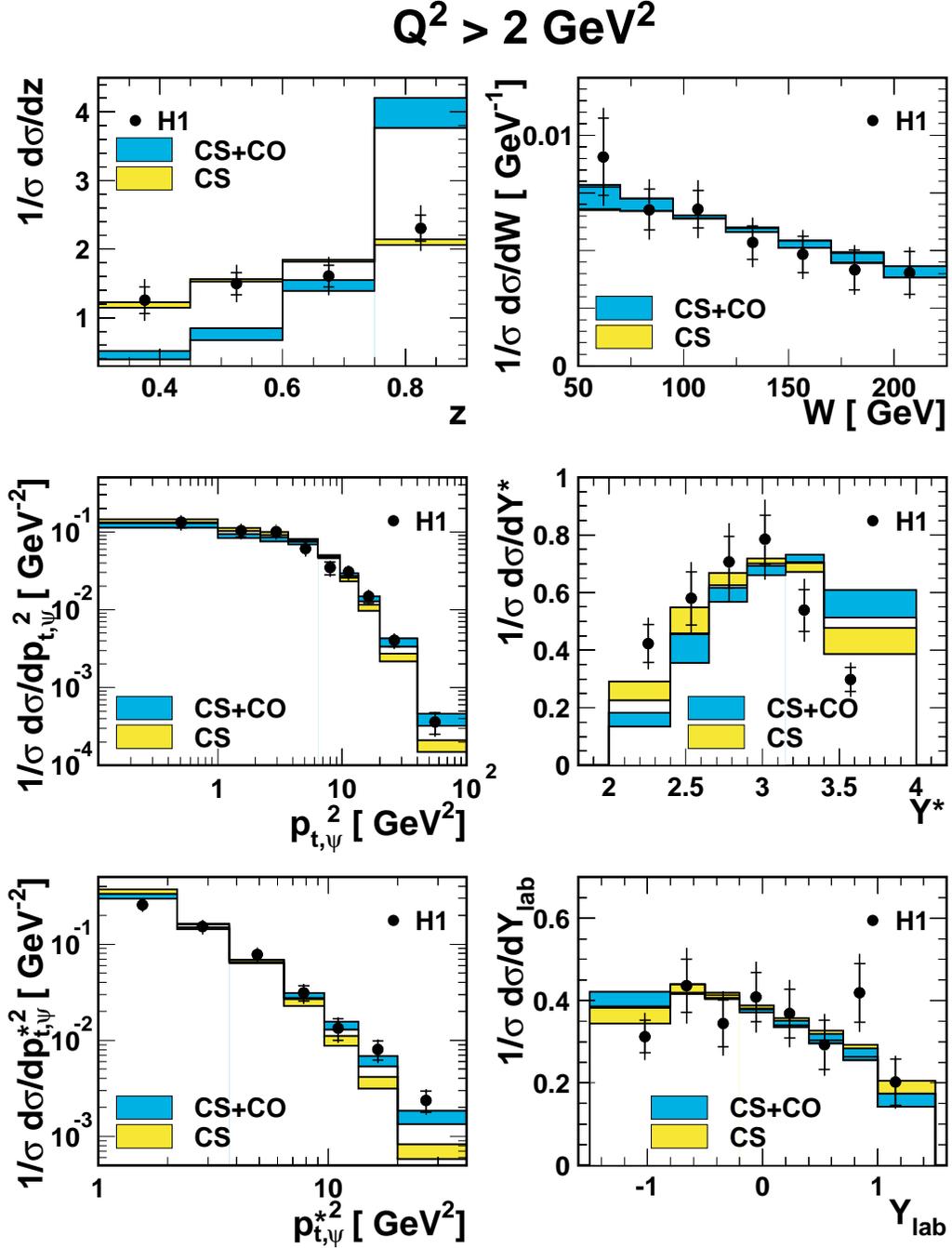,height=18cm,bbllx=44pt,bblly=108pt,bburx=527pt,%
bbury=747pt,clip=}
}
\vspace*{-1.cm}
\caption{Comparison of the CSM and NRQCD predictions \protect\cite{ep} for the
normalized $z$, $W$, $p_{t,\psi}^2$, $Y^\star$, $p_{t,\psi}^{\star2}$, and
$Y_{\rm lab}$ distributions of $e+p\to e+J/\psi+X$ with H1 data
\protect\cite{h1}.}
\label{fig:h1n}
\end{figure*}

In Figs.~\ref{fig:h1} and \ref{fig:h1n}, our CSM and NRQCD predictions
\cite{ep} are confronted with data from the H1 Collaboration at HERA
\cite{h1}.
Here, $p_{t,\psi}$ and $Y$ are the $J/\psi$ transverse momentum and rapidity,
respectively, $W$ is the $\gamma^\star p$ invariant mass, and quantities
referring to the $\gamma^\star p$ centre-of-mass frame are denoted by an
asterisk.
We take the charm-quark mass to be $m_c=1.5$~GeV, employ the LO proton PDFs
from Ref.~\cite{mrst}, with $\Lambda^{(3)}=204$~MeV, adopt the corresponding
MEs from Ref.~\cite{bkl}, and choose the renormalization and factorization
scales to be $\mu=M=\sqrt{Q^2+4m_c^2}$.
The theoretical uncertainties are conservatively estimated by taking into 
account the experimental errors on $m_c$ and the MEs and the well-known
ambiguity between
$\left\langle{\cal O}^{J/\psi}\left[{}^1\!S_0^{(8)}\right]\right\rangle$ and
$\left\langle{\cal O}^{J/\psi}\left[{}^3\!P_0^{(8)}\right]\right\rangle$
\cite{ano}, using alternative proton PDFs, and varying the scales.
As for the normalization, the H1 data clearly favours the NRQCD prediction
(see Fig.~\ref{fig:h1}).
On the other hand, the shapes of the various distributions exhibit a diverse
pattern (see Fig.~\ref{fig:h1n}).
While the H1 measurement nicely agrees with NRQCD in the normalized
$p_{t,\psi}^2$ and $p_{t,\psi}^{\star2}$ distributions, it favours the
normalized $z$ distribution of the CSM, which distinctly undershoots the NRQCD
prediction in the upper $z$ range.
The latter feature is familiar from inclusive $J/\psi$ photoproduction at HERA
\cite{ano}.

\boldmath
\section{$\nu N$ DIS via the weak neutral current}
\unboldmath

\vspace*{-1.cm}
\begin{figure}[h]
\epsfig{figure=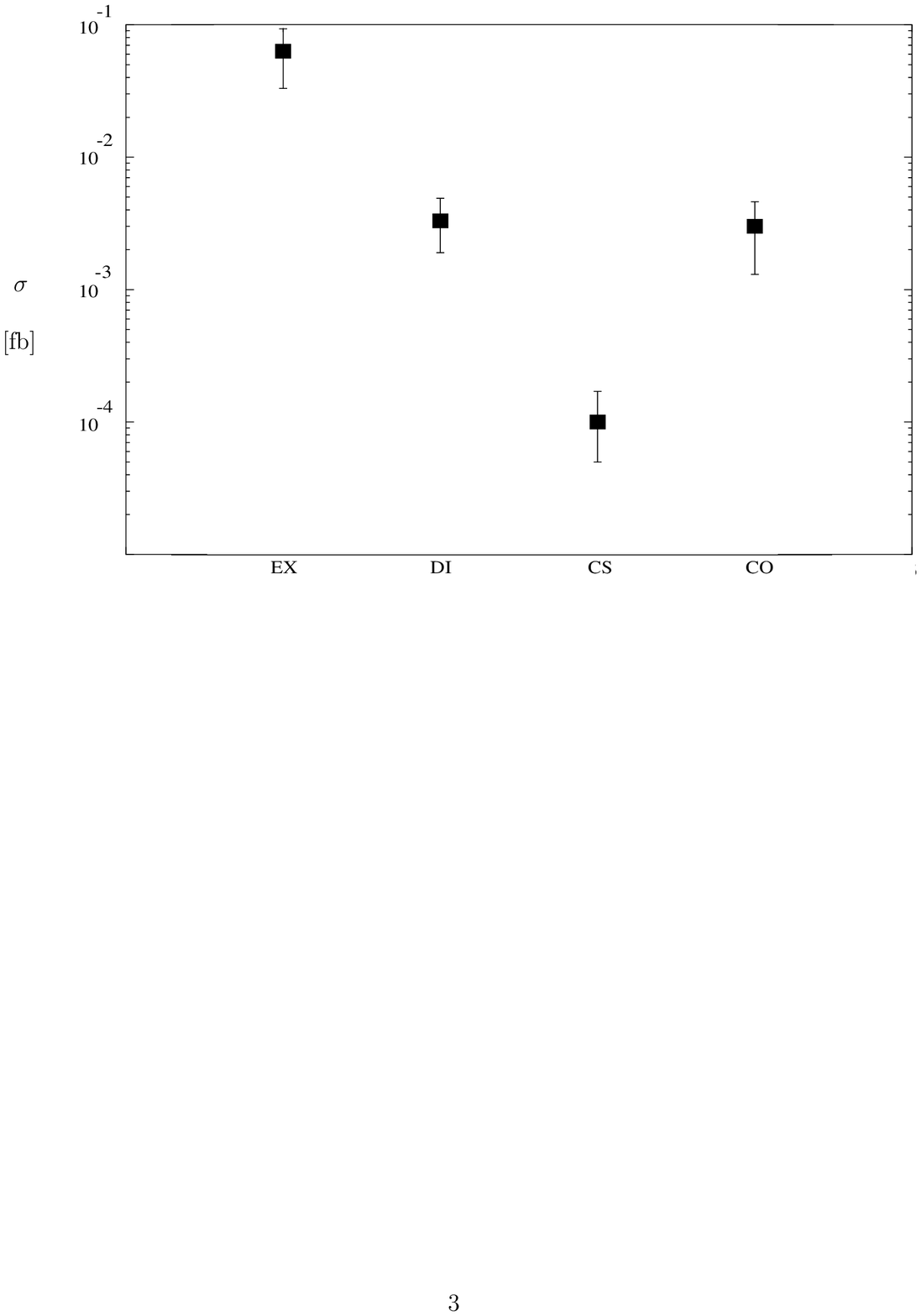,width=7.5cm,bbllx=75pt,bblly=439pt,bburx=527pt,%
bbury=720pt,clip=}
\vspace*{-1.cm}
\caption{The total cross section of $\nu+{\rm Pb}\to\nu+J/\psi+X$ measured by
CHORUS \protect\cite{chorus} (EX) is compared with the predicted
\protect\cite{nun} diffractive (DI), CS, and CO contributions.}
\label{fig:nun}
\end{figure}
\vspace*{-0.75cm}
The total cross section of $\nu+N\to\nu+J/\psi+X$ in NC DIS was recently 
measured by the CHORUS Collaboration \cite{chorus}, who exposed a lead target
in the CERN wide-band neutrino beam.
No acceptance cuts were imposed to exclude the $Z^\star g$-fusion and
diffractive processes, so that their contributions have to be included in
the theoretical prediction.
Furthermore, no selection of directly produced $J/\psi$ mesons was
implemented, and those from $\chi_{cJ}$ and $\psi^\prime$ decays need to be
considered, too.
However, the fraction of $J/\psi$ mesons from $B$ decays should be negligible 
\cite{nun}.
In Fig.~\ref{fig:nun}, the CHORUS result, $(6.3\pm3.0)\times10^{-2}$~fb, is
compared with our predictions \cite{nun} for the diffractive, CS and CO
contributions.
The diffractive contribution is estimated by means of the vector-meson 
dominance model as in Ref.~\cite{kue}.
The combined prediction undershoots the CHORUS central value by almost one
order of magnitude, but the experimental error is still rather sizeable.

\boldmath
\section{$ep$ and $\nu N$ DIS via the charged current}
\unboldmath

\begin{figure}[h]
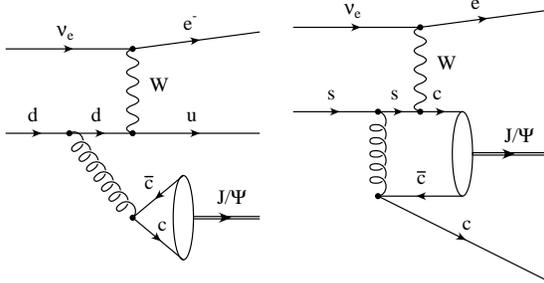

\begin{tabular}{cc}
\parbox{3.4cm}{\epsfig{file=CC1.epsi,width=3.4cm}}
&
\parbox{3.4cm}{\epsfig{file=CC2.epsi,width=3.4cm}}
\end{tabular}
\vspace*{-1.cm}
\caption{Feynman diagrams for $\nu+N\to l^\pm+J/\psi+X$ in DIS.}
\label{fig:nunfey}
\end{figure}

The reaction $\nu+N\to l^\pm+J/\psi+X$ proceeds through the Feynman diagrams 
depicted in Fig.~\ref{fig:nunfey}.
If the hadronic remnant $X$ is charmless, then we are dealing with a CO 
process.
Otherwise, both CS and CO channels contribute.
Since the $W$ boson cannot fluctuate into charmonia, there are no diffractive
processes.
For the total cross section of prompt $J/\psi$ production under CHORUS 
experimental conditions, we find
$4.9{+3.4\atop-2.3}\times10^{-5}$~fb in the charmless-$X$ mode and
$1.8{+1.5\atop-0.9}\times10^{-5}$~fb in the charmed-$X$ mode.


\vspace*{-1.2cm}

\section{Conclusions}

Inclusive $J/\psi$ production in DIS lends itself as a sensitive probe of the
CO mechanism.
As for $e+p\to e+J/\psi+X$, the H1 data \cite{h1} generally confirms NRQCD and
disfavours the CSM \cite{ep}.
However, NRQCD predicts at LO a distinct rise in cross section as $z\to1$,
which is not reflected by the H1 data.
This anomaly is familiar from photoproduction, and it is likely to be resolved
by the inclusion of higher-order corrections \cite{ano}, possibly in 
combination with intrinsic-$k_T$ effects and/or nonperturbative shape
functions.
As for $\nu+N\to\nu+J/\psi+X$, the CHORUS central value for the total cross
section \cite{chorus} exceeds the LO prediction \cite{nun} by almost one order
of magnitude.
However, the experimental error is still rather sizeable.
As for $\nu+N\to l^\pm+J/\psi+X$, CO processes are dominant, diffractive ones
are absent, and the experimental signature is spectacular, so that a
measurement would be worthwhile \cite{cc}.
Inclusive $J/\psi$ production in CC DIS represents a challenge for HERA and
THERA.

\end{document}